\documentclass[nomathfonts]{aipproc}
\layoutstyle{6s}
\usepackage{xspace,epic,eepic,graphicx}
\usepackage{epsfig}
\usepackage{natbib}
\graphicspath{{figures/}}

\title[Bayesian blind component separation for CMB observations.]
      {Bayesian blind component separation for cosmic microwave 
background observations.}
\author{H. Snoussi }
  {
  address = {Laboratoire des Signaux et Syst{\`e}mes (L2S), 
\linebreak   
             Sup{\'e}lec, Plateau de Moulon,  
                                 91192 Gif-sur-Yvette Cedex, France}
  ,email   = snoussi@lss.supelec.fr
  }
\author{G. Patanchon}
{
 address = {PCC -- Coll{\`e}ge de France, \linebreak
            11, Place Marcelin Berthelot, 
                F-75231 Paris, France},
 email =  patanchon@cdf.in2p3.fr
}
\author{J.F. Mac{\'\i}as-P{\'e}rez}
{
 address = {PCC -- Coll{\`e}ge de France, \linebreak
            11, Place Marcelin Berthelot, 
                F-75231 Paris, France},
 email =  macias@cdf.in2p3.fr
}

\author{A. Mohammad-Djafari}
  {
  address = {Laboratoire des Signaux et Syst{\`e}mes (L2S), \linebreak  
             Sup{\'e}lec, Plateau de Moulon, 
                                 91192 Gif-sur-Yvette Cedex, France}
  ,email   = djafari@lss.supelec.fr
  }
\author{J. Delabrouille}
{
 address = {PCC -- Coll{\`e}ge de France, \linebreak
            11, Place Marcelin Berthelot, 
                F-75231 Paris, France},
 email =  delabrouille@cdf.in2p3.fr
}

\input{alphabet}       % Definitions de macros pour les alphabets
\input{abrege}         % Abbreviations latines, anglaises et fran‡aises
\def\cro#1{\left[#1\right]}

\def\bigcro#1{\bigl[#1\bigr]}

                \def\Tr#1{{\mathrm{tr}}\bigcro{#1}}                     
                                     
\def\esp{{\mathrm{E}}\,}                \def\Esp#1{{\mathrm{E}}\bigcro{#1}}

\def\Exp#1{\exp\cro{#1}}                %\def\Expold#1{\exp\cro{#1}}
\newsavebox{\fminibox}
\newlength{\fminilength}
\newenvironment{fminipage}[1][\linewidth]
  {\setlength{\fminilength}{#1}%-2\fboxsep-2\fboxrule}%
   \begin{lrbox}{\fminibox}\begin{minipage}{\fminilength}}
  {\end{minipage}\end{lrbox}\noindent\fbox{\usebox{\fminibox}}}

  \def\+{^\dagger}
\def\nequiv{\not\kern-.05em\equiv}
\def\egal{\kern-.5em=\kern-.5em}        % Moins d'espace autour de "="
\def\propt{\kern-.2em\propto\kern-.2em} % Idem
\def\wh#1{\widehat{#1}}                 % Sombrero !

\def\argmax{\mathop{\mathrm{arg\,max}}} % Mieux que \def\argmax{\arg\max}
\def\intdouble{\int\kern-0.3em\int}
\def\inttriple{\int\kern-0.3em\int\kern-0.3em\int}

\def\rond#1{\overset{\kern-0.33em~_\circ}{#1}}
\def\rondit[#1]#2{\overset{\kern#1~_\circ}{#2}}

\def\babs{\begin{abstract}}             \def\eabs{\end{abstract}}
\def\barr{\begin{array}}                \def\earr{\end{array}}
\def\bcc{\begin{center}}                \def\ecc{\end{center}}

\def\bdes{\begin{description}}          \def\edes{\end{description}}
\def\bdoc{
\begin{document}}             \def\edoc{\end{document}}
\def\ben{\begin{enumerate}}             \def\een{\end{enumerate}}
\def\beqn{\begin{eqnarray}}             \def\eeqn{\end{eqnarray}}
\def\beqnl#1{\beqn\label{#1}}           \def\eeqnl#1{\label{#1}\eeqn}
\def\beqnx{\begin{eqnarray*}}           \def\eeqnx{\end{eqnarray*}}
\def\bseqn{\begin{subeqnarray}}         \def\eseqn{\end{subeqnarray}}
\def\beq#1\eeq{\begin{equation}#1\end{equation}}
\def\bal#1\eal{\begin{align}#1\end{align}}
\def\balx#1\ealx{\begin{align*}#1\end{align*}}
\def\beqx{$$}                           \def\eeqx{$$}
\def\bfig{\protect\begin{figure}}       \def\efig{\protect\end{figure}}
\def\bfigx{\protect\begin{figure*}}     \def\efigx{\protect\end{figure*}}
\def\bfigt{\protect\begin{figurette}}   \def\efigt{\protect\end{figurette}}
\def\bfl{\begin{flushleft}}             \def\efl{\end{flushleft}}
\def\bfr{\begin{flushright}}            \def\efr{\end{flushright}}
\def\bit{\begin{itemize}}               \def\eit{\end{itemize}}
\def\bmi{\begin{minipage}}              \def\emi{\end{minipage}}
\def\bfmi{\begin{fminipage}}            \def\efmi{\end{fminipage}}
\def\bpic{\begin{picture}}              \def\epic{\end{picture}}
\def\bqu{\begin{quote}}                 \def\equ{\end{quote}}
\def\bqun{\begin{quotation}}            \def\equn{\end{quotation}}
\def\bsl{\begin{slide}}                 \def\esl{\end{slide}}
\def\btabb{\begin{tabbing}}             \def\etabb{\end{tabbing}}
\def\btabl{\begin{table}}               \def\etabl{\end{table}}
\def\btablx{\begin{table*}}             \def\etablx{\end{table*}}
\def\btab{\begin{tabular}} %\def\etab{\end{tabular}} Conflit avec eta gras... Elle est pas grasse, ma soeur !
\def\btabu{\begin{tabular}}             \def\etabu{\end{tabular}}
\def\btabx{\begin{tabular*}}            \def\etabx{\end{tabular*}}
\def\bbib{\begin{thebibliography}{999}} \def\ebib{\end{thebibliography}}
\def\bver{\begin{verbatim}}             \def\ever{\end{verbatim}}
\def\bca{\begin{cases}}                          \def\eca{\end{cases}}
       % Abbreviations des \begin{} et des \end{}

%----------- MACROS et Definitions Speciales pour CE document ----------%%
\def\pxta{p\left( \xb(1),\ldots,\xb(T) | \Ab \right)}
\def\pxas{p\left( \xb_{1..T} | \Ab, \sb_{1..T} \right)}
\def\pasx{p\left( \Ab,\sb_{1..T} | \xb_{1..T} \right)}
\def\pax{p\left( \Ab | \xb_{1..T} \right)}
\def\psx{p\left( \sb_{1..T} | \xb_{1..T} \right)}
\def\sbh{\wh{\sb}}
\def\abh{\wh{\ab}}
\def\Sbh{\wh{\Sb}}
\def\Abh{\wh{\Ab}}
\def\Bbh{\wh{\Bb}}
\def\zbh{\wh{\zb}}
\def\xbh{\wh{\xb}}
\def\Reps{\Rb_{\epsilon}}
\def\Rz{\Rb_{z}}
\def\muz{\mub_{z}}
\def\pz{p_{z}}
\def\thetaz{\thetab_{z}}
\def\Gz{\Gammab_{z}}
\def\Tr#1{\mbox{Tr}\left( #1 \right)}

% MACROS et definitions pour astro bib.
%include bibtex definitions
\def\mnras{MNRAS}
\def\nature{Nat.} 
\def\apj{ApJ}
\def\apjl{ApJ}
\def\aap{A\&A}
\def\aaps{A\&A}
\def\prd{Phys. Rev. D}

\bdoc
     
\begin{abstract}
We present a technique based on the Expectation-Maximization (EM) 
algorithm for the separation of the components of noisy mixtures in 
the Fourier plane.  We perform a semi-blind joint estimation of 
components, mixing coefficients and noise rms levels.  
A priori information for the spatial spectrum of the components and
for the mixing coefficients can be naturally included in the 
algorithm.
This method 
is applied to the separation of distinct astrophysical emissions  
on simulations of future observations with 
the High Frequency Instrument of the Planck space mission, due to be 
launched in 2007.  The simulations include a mixture of astrophysical 
emissions and instrumental white noise at the levels expected for 
this 
instrument.  We have obtained good preliminary results with this 
technique, being able to blindly separate noisy mixtures with 3 
components.

\end{abstract}

\maketitle
\thispagestyle{empty}

\section{Introduction}

The restitution of signals or images from the observation of their 
mixtures has grown into a field of itself now classically called 
``source separation". Astrophysics, being a field of physics in which 
nearly all the information we can get about the physical processes occurring
in very distant places is through observation of their electromagnetic emission,
is naturally a field in which source
separation methods can be usefully applied. One such application, of 
particular importance, can be found in millimeter and submillimeter 
astronomy.

Mapping and interpreting sky emissions in the millimeter and 
sub-millimeter range recently made possible thanks to dedicated sensitive
balloon borne and space borne instruments, is indeed one of the main objectives
of present and 
upcoming observational effort in astronomy.  Among 
the scientific objectives of these observations, the precise measurement 
of primordial temperature and/or polarisation fluctuations of the 
Cosmic Microwave Background (CMB) radiation is one of the 
priorities, which has been given recently a tremendous emphasis.  
This radiation, emitted some 12-15 billion years ago, conveys a 
large amount of information about our universe as a whole.  
% Observations with the COBE satellite in 1992 
% \cite{1992ApJ...396L...1S} provided the first low angular resolution 
% ($\sim 7^{\circ}$) full sky map of the spatial fluctuations of CMB 
% emission (traditionally called anisotropies) and the normalization of 
% their power spectrum.  This first detection was followed by many 
% more.  
The importance of measuring anisotropies of the Cosmic Microwave 
Background (CMB) to constrain cosmological models is now well 
established.  In the past ten years, tremendous theoretical activity 
demonstrated that measuring the properties of these temperature 
anisotropies will constrain drastically the cosmological parameters 
describing the matter content, the geometry, and the evolution of our 
Universe \cite{1996ApJ...471..542H,1996PhRvL..76.1007J}.
Recently, balloon-borne experiments such as Boomerang 
\cite{2000Natur.404..955D} and MAXIMA \cite{2000ApJ...545L...5H} have 
measured the CMB anisotropies in small patches of the sky at higher 
angular resolution ($\sim 0.2 ^{\circ}$) placing strong constraints 
on the quasi-flatness of the Universe.
% and on the unexpected high value 
% of 
% the cosmological constant, $\Lambda$.
A new generation of satellite experiments 
will provide shortly multi-frequency observations of the microwave 
and far infrared emission of the sky, with as a main objective the 
precise mapping of CMB fluctuations over the sky at high angular 
resolution and with unprecedented accuracy. One of these missions, 
the Microwave Anisotropy Probe, has been launched by NASA end of june 
2001, and will provide full sky 15-30 arcminute resolution maps of the sky in 
three frequency channels with high signal to noise ratio in each 
pixel. Even more sensitive by an order of magnitude, the Planck 
mission, to be launched by ESA in 2007, will provide full sky maps 
with 5-30 arcminute resolution in 9 frequency channels between 30 and 
850 GHz.

The accuracies required for precision tests of the cosmological 
models, however, is such that it is necessary to achieve precisions on 
the CMB maps well below the expected level of contamination from 
astrophysical ``foregrounds". Indeed, there are at least 6 different physical
emission processes which will contribute significant components in the Planck
observations. Thus, it is crucial for the success of 
these future missions to separate CMB and foregrounds in the observed 
microwave maps.  The separation of these emissions by adapted source 
separation methods is expected to be one of the main steps in the 
analysis of future CMB data.

So far, two sets of independent algorithms have been proposed: MEM and 
Wiener filtering \cite{1999NewA....4..443B,Tegmark96,1998MNRAS.300....1H}
for which the 
electromagnetic spectrum of the components is assumed known, and blind 
Independent Components Analysis (ICA) \cite{2000MNRAS.318..769B} for 
which no a priori is assumed.  The former algorithms give promising 
results although are strongly limited by the uncertainties in the 
electromagnetic spectrum of the components which, as we indicated 
above, can be severe for some of them.  The ICA algorithm has shown 
promising results for simplified non noisy mixtures but has not yet 
attained a sufficient grade of sophistication to account for 
instrumental noise and beam smoothing.  \\

We propose an alternative method for the separation of components in 
multi-frequency CMB data, based on the exploitation of the spectral 
diversity of the data.  The maximisation of the likelihood is achieved 
with an Expectation-Maximization (EM) algorithm.  Our method permits 
the simultaneous estimation of the spatial distribution of the 
components and of their electromagnetic continuum spectrum of 
emission.  In section 2 we describe the basic model for noisy mixtures 
in the framework of the separation of CMB and foregrounds.  In 
section 3 we present simulations of the HFI Planck observations which 
are used to test the separation algorithm.  Section 4 describes in 
detail the EM algorithm applied to the separation of components.  
Section 5 summarizes the main results we obtain by applying the EM 
algorithm to our simulations.
 
\section{Modeling CMB data and foregrounds}

We classify the main relevant astrophysical components in the 
millimetre range in three kinds of components. The CMB anisotropy signal,
cosmological in 
origin, has been emitted in the very distant past as a relic 
radiation from times when the universe was fully ionised and before 
astrophysical objects as galaxies and clusters formed. Extra-galactic 
foregrounds, less distant in origin, are due to emissions coming from 
outside our galaxy. Galactic components, finally, originate from our 
own galaxy, and are strongly peaked towards the galactic plane. The 
main emissions at millimeter wavelengths can be summarised as:

\begin{enumerate}
\item{{\bf CMB anisotropies.}}
\item{{\bf Extra-Galactic Foregrounds}
\begin{itemize}
\item Point sources (radio-galaxies, infrared galaxies, quasars).
\item The Sunyaev-Zeldovich (SZ) emission in clusters of galaxies.
%\item The Cosmic Infrared Background (CIB).
\end{itemize}
}
\item{{\bf Galactic Foregrounds}
\begin{itemize}
\item Dust emission: thermal emission from intragalactic cold dust 
grains.
\item Synchrotron emission: radiation from relativistic electrons in 
Galactic magnetic fields.
\item Free-Free (Bremsstrahlung) emission: radiation from free 
Galactic electrons. 
\end{itemize}
}
\end{enumerate}

These components are known to have different spectral emission laws as 
a function of the observing frequency $\nu$.  Therefore, the 
separation of the various emissions can be achieved using 
multi-frequency observations, i.e. the observation of the sky at 
different wavelengths, with component separation techniques based on 
the diversity (and possibly the prior knowledge) of electromagnetic 
spectra of foregrounds and CMB, and also on the spatial statistical 
independence of the different components.  For the CMB and SZ effect 
the electromagnetic spectrum is accurately known and can be included 
in the separation methods \cite{1998MNRAS.300....1H}.  However, for 
the rest of the components we dispose, in the best of the cases, only 
of spectra extrapolated from distant frequencies 
\cite{1999tkc..conf..204D}.  The spatial spectrum of the components is 
not known although reasonably good extrapolations can be obtained from 
observed data at lower resolution \cite{1999NewA....4..443B}.  \\

As a first step in discussing component separation techniques, we 
present the basic model which can be used to describe the observed sky 
emission $y_{\nu} (\rb)$ at position $\rb$ and at frequency $\nu$.  In 
the millimeter and centimeter range of the electromagnetic spectrum, 
$y_{\nu} (\rb)$ can be considered as a linear superposition of CMB 
radiation $\hat{s}_{CMB} (\nu,\rb)$ and foreground emissions 
$\hat{s}_{f}(\nu,\rb)$ convolved with the instrumental response of the 
detector, $b_{\nu}(\rb)$, which is here assumed to be symmetric for simplicity.
We have :
\beq
\label{modelfreqpos}
y_{\nu} (\rb) = \hat{s}_{CMB} (\nu,\rb)*b_{\nu}(\rb) + 
\sum_{f=1}^{N_{f}} \hat{s}_{f}(\nu,\rb)*b_{\nu}(\rb) + n_{\nu} (\rb) 
\eeq 
$N_{f}$ represents the number of independent foreground 
components considered, $*$ defines the convolution operator and 
$n_{\nu} (\rb)$ is the instrumental noise of the detector.\\

In the case of the CMB radiation the spatial and electromagnetic 
frequency dependence can be separated,
$$
\hat{s}_{CMB} (\nu,\rb) = g_{CMB}(\nu) \times s_{CMB} (\rb) 
$$
being $g_{CMB}(\nu)$ and $s_{CMB} (\rb)$ the CMB electromagnetic 
spectrum and spatial distribution respectively.  For most foreground 
components we can also, at least to first approximation, separate the 
spatial and frequency terms 
\cite{1999NewA....4..443B,1998MNRAS.300....1H} so that equation 
\ref{modelfreqpos} reads
$$
 y_{\nu} (\rb) = g_{CMB} (\nu) \; s_{CMB} (\rb) *b_{\nu}(\rb)
+ \sum_{f=1}^{N_{f}} g_{f}(\nu) \; s_{f} (\rb)*b_{\nu}(\rb)
+ n_{\nu} (\rb)
$$
where $g_{f}$ represents the mean electromagnetic spectrum of the 
foreground components. Typical spatial templates for CMB, dust and SZ 
emissions are shown in figure \ref{compspdist}, and the spectral 
emission law (as a function of electromagnetic frequency) in figure
\ref{compelecspc}.\\

%%%%%%%%%%%%%%%%%%%%%%%%%%%%%%%%%%%%%%%%%%%%%%%%%%%%%%%%%%%%%%%%%%%%%%
\begin{figure}
\label{compspdist}
\includegraphics[scale=0.73,draft=false]{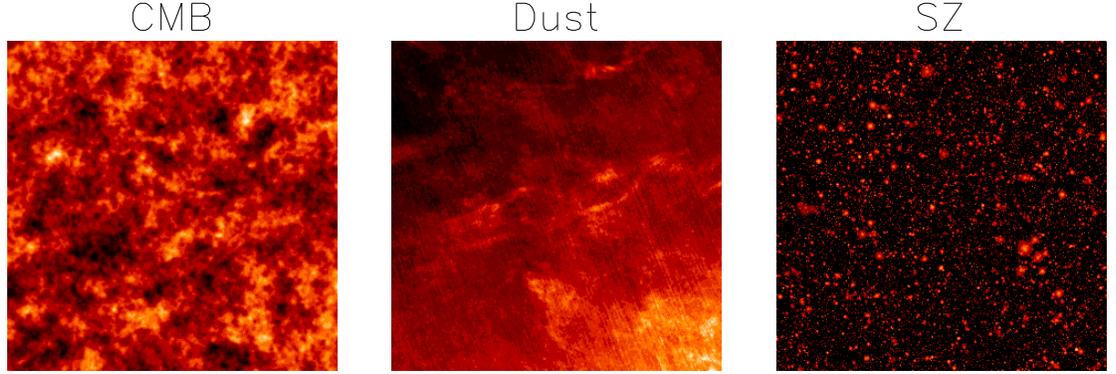}
\caption{Spatial template of the CMB, dust and SZ components used
in the simulations presented in the text. For visibility the SZ
template is displayed in log scale.}
\end{figure}
%%%%%%%%%%%%%%%%%%%%%%%%%%%%%%%%%%%%%%%%%%%%%%%%%%%%%%%%%%%%%%%%%%%%%%

%%%%%%%%%%%%%%%%%%%%%%%%%%%%%%%%%%%%%%%%%%%%%%%%%%%%%%%%%%%%%%%%%%%%%%
\begin{figure}
\label{compelecspc}
\includegraphics[scale=0.45,draft=false]{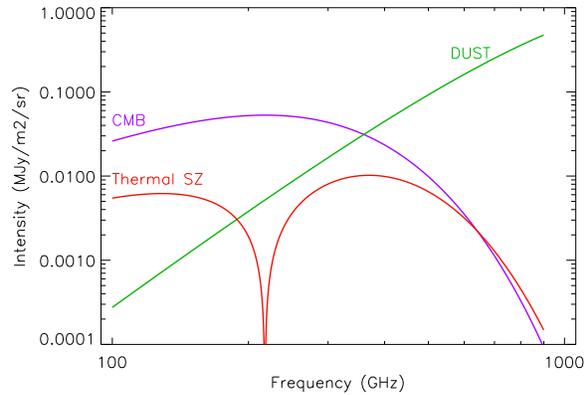}
\caption{Electromagnetic spectrum of the CMB, dust and SZ components 
used
in the simulations presented in the text. These spectra fully define 
the missing matrix
$Ab$ when the beam smoothing effects are neglected.}
\end{figure}
%%%%%%%%%%%%%%%%%%%%%%%%%%%%%%%%%%%%%%%%%%%%%%%%%%%%%%%%%%%%%%%%%%%%%%

From the modeling point of view, the CMB component is not special and 
therefore we can write
$$
 y_{\nu} (\nu,\rb) = \sum_{i=1}^{N_{c}} g_{i}(\nu) \; 
 s_{i}(\rb)*b_{\nu}(\rb)+ n_{\nu} (\rb)
$$
where $N_{c} = N_{f}+1$.  Thus, for a set of detectors 
$d={1,..,N_{d}}$ operating at electromagnetic frequencies $\nu_{d}$ 
\beq
\label{rmixingeq}
y_{d}(\rb) = \sum_{i=1}^{N_c} A_{d i} \; s_{i}(\rb)*b_{d}(\rb)+ 
n_{d}(\rb) \eeq where $A_{d i} = g_{i}(\nu_{d})$ is a $N_{d} \times 
N_{c}$ matrix called the mixing matrix.  \\

If we assume the instrumental response of the detectors to be 
spatially invariant it is more convenient to work in Fourier space, 
since the convolution in equation \ref{rmixingeq} becomes a simple 
multiplication and we obtain 
\beq
\label{fmixingeq}
\tilde{y}_{d}(\kb) = \sum_{i=1}^{N_c} \tilde{A}_{d i}(\kb) \; 
\tilde{s}_{i}(\kb) + \tilde{n}_{d}(\kb) \eeq where $\tilde{A}_{d 
i}(\kb) = A_{d i} \tilde{b}_{d}(\kb)$.  Further, equation 
\ref{fmixingeq} is verified for each Fourier mode $\kb$ independently 
and if the components are assumed to be stationary random variables 
then the separation can be performed also independently at each 
Fourier mode.

Hereafter, to simplify the mathematical formalism in the following 
sections, we will not
include the beam smoothing in the analysis, so that equation 
\ref{fmixingeq} in
matrix notation and for each mode $\kb$ becomes
\beq
\label{matmixingeq}
\yb (\kb) = \Ab \; \sb (\kb) + \nb (\kb) 
\eeq 
where $\yb$, $\sb$ and $\nb$ are column vectors containing $N_{d}$, 
$N_{c}$ and $N_{d}$ complex components respectively, and the matrix 
$\Ab$ has dimensions $N_{d} \times N_{c}$.

Note that the separation of components requires the joint estimation 
of the
sources $\sb$ and the mixing matrix $\Ab$ (or at least some of its elements).
In practice, the noise 
covariance
matrix, $\Rb_{\epsilon}= E\left(\nb \nb^{T} \right)$ is not very well 
known and in most cases has also
to be estimated from the observed data, $\yb$.

\section{Simulated Observations for the Planck High Frequency Instrument}

The Planck satellite will measure the emission from the whole sky in 
the electromagnetic frequency range 30 to 850 GHz.  The data will be 
taken using two independent instruments, the Low Frequency Instrument 
(LFI) and the High Frequency Instrument (HFI).  Here, we only consider 
(as a starting point for devising and testing component separation 
methods) the HFI instrument which observes the sky in 6 frequency channels 
between 100 and 850 GHz. In the frequency range of the HFI 
instrument, the CMB radiation dominates at low frequencies and the 
dust emission at high frequencies.  In addition, we expect significant 
contribution from the SZ effect in clusters of galaxies.  The rest of 
the foreground components present weak emission at the HFI 
electromagnetic frequencies.  \\

We simulate, following the model introduced in section 2, observations 
from 6 detectors ($N_{d}=6$) corresponding to the 6 HFI frequency 
channels.  To reduce the computing time we restrict our simulations to 
small patches of the sky of $300 \times 300$ square pixels of 2.5 
arcmin as described in \cite{jgpaper}.  Working on small maps finds 
also a theoretical justification in the fact that the spectrum of 
emission of the dust (for instance) is known to vary slightly from a 
region of the sky to the other. The instrumental noise for 
each detector is considered white and Gaussian of zero mean and of rms 
expected for the HFI instrument channels.  We also assume no 
correlation between the noise of the different detectors.  \\

The simulations include 3 components ($N_{c}=3$), CMB radiation, dust 
emission and SZ effect in clusters of galaxies.  The CMB component is 
a Gaussian randomly generated realisation of CMB anisotropies obtained 
from a theoretical spatial power spectrum calculated with CMBFAST 
\cite{cmbfastpaper}.  The electromagnetic frequency dependence of the 
CMB component is the derivative with respect to temperature of a 
Planck Blackbody spectrum, at temperature T=2.726 K. The dust 
component was obtained by extrapolation of the IRAS satellite 100 
$\mu$m map (which serves as a spatial template) to the HFI 
frequencies, with an electromagnetic spectrum proportional to $\nu^2 
B_{\nu}(T=17.5\, {\rm K})$.  Finally, the spatial template for the SZ 
component, due to the inverse-Compton scattering of primordial photons 
by hot electrons in the ionised gaz present in clusters of galaxies, 
was obtained from a simulation of the spatial distribution of the SZ 
comptonization parameter \cite{jjpaper}. The emission law of the SZ 
emission is known theoretically, and depends only on the frequency of 
observation (in the Kompaneets approximation, which we assume here, 
see \cite{szpaper} for details).\\
The simulated observations are obtained from the superposition of the 
three above spatial templates with mixing coefficients set by the 
electromagnetic spectra of emission and the frequency of observation.  
The observations used as inputs in the separation procedure (with 
noise added at the level expected for Planck HFI detectors) are shown 
in figure \ref{simobs}.

%%%%%%%%%%%%%%%%%%%%%%%%%%%%%%%%%%%%%%%%%%%%%%%%%%%%%%%%%%%%%%%%%%%%%
\begin{figure}
\label{simobs}                      
\includegraphics[scale=0.73,draft=false]{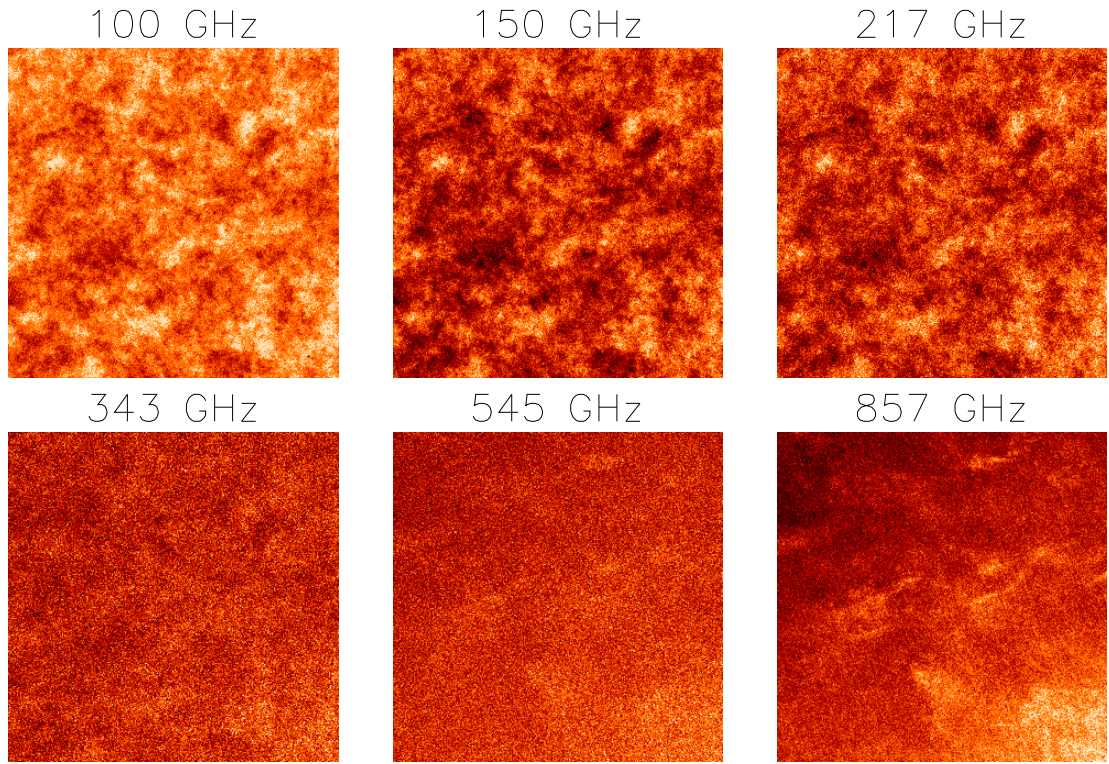}
\caption{Simulated observations of the six HFI channels.}
\end{figure}
%%%%%%%%%%%%%%%%%%%%%%%%%%%%%%%%%%%%%%%%%%%%%%%%%%%%%%%%%%%%%%%%%%%%%

\section{The separation method}

The blind separation of sources constitutes a common example of {\it missing 
data problem} (problems for which only partial information is 
available) in the literature \cite{EMbook}.  Indeed, following 
equation \ref{matmixingeq} and considering the estimation of the 
mixing matrix, $\Ab$, from a set of observations $\yb$, we can 
consider the components of the mixture, $\sb$ as the missing data in 
our problem because they are not directly available to the observer.  
As discussed in the previous sections, we are also interested in the 
estimation of the noise covariance matrix, $\Rb$, which can be 
considered as an additional parameter in the separation.

Given the observations $\yb$, our approach is thus to estimate first 
the mixing 
matrix $\Ab$ and the noise covariance matrix $\Reps$ by maximizing 
the log-likelihood function $\log\,{\bf 
p}(\yb_{1..K}\,|\,\Ab,\,\Reps)$,
\[
(\wh{\Ab},\,\wh{\Rb}_{\epsilon})=\argmax_{\Ab,\,\Reps}\; {\log\,{\bf 
p}(\yb_{1..K}\,|\,\Ab,\,\Reps)}
\]
The maximization of this log-likelihood will be made with a specific 
implementation of the Expectation Maximization
(EM) algorithm introduced by Dempster-Laird-Rubin in 
\cite{Dempster77}. The estimation of the source templates
 is done afterwards by inverting the linear system $y=As+n$ using a 
classical inversion method as done by Bouchet and Gispert in
\cite{1999NewA....4..443B}.

\subsection{The EM algorithm: formalism}

Let's consider a missing data problem with observed data $(\yb(k))_{k 
\in \{1..K\}}$ and missing data $(\xb(k))_{k \in \{1..K\}}$, for which 
we want to estimate the set of parameters, $\thetab$.  Hereafter, we 
will note as the incomplete-data problem that for which only the 
observed data are used to obtain the unknown parameters, and the 
complete-data problem, that for which the missing data are also 
considered in the solution.

An optimal estimate of $\thetab$, $\hat{\thetab}$ can be obtained by 
maximizing its incomplete log-likelihood function,
$$
\mathcal{L}(\thetab) = \log {\bf p}(\yb\,|\,\thetab)
$$ 
However, in many practical cases, this function is complex and 
difficult to work with and it is more convenient to solve the 
complete-data problem.  \\

The EM algorithm permits to solve the incomplete data problem.  It is 
an iterative algorithm which generates a sequence of approximations to 
find the maximum observed likelihood estimator when only partial 
information is available, by marginalizing at each iteration over the 
missing data.  \\

At iteration $j+1$ of the algorithm we can write
\beq\label{transf}
\thetab^{j+1}=\Mc(\thetab^{j})
\eeq
where  $\Mc$ is a mapping function, named the re-estimation 
transformation. After the initialization to 
an arbitrary point $\thetab^{0} \in \Thetab$, the new estimate of 
$\thetab$ is computed using equation (\ref{transf}),
until a fixed point is obtained such that 
$\thetab^{j+1}=\thetab^{j}$. The mapping $\Mc$ is performed in two 
steps
\bit
\item {\bf $\esp$-step}: Computation of  $Q(\thetab, 
\thetab^{j})=\esp\;\left[\log{{\bf p}(\yb, \xb\,|\,\thetab)}\,|\,\yb, 
\thetab^{j}\right] $ 
\item {\bf M-step}: Find 
$\thetab^{j+1}=\displaystyle{\argmax_{\thetab \in 
\Thetab}\;{Q(\thetab, \thetab^{j})}}$
\eit
where ${\bf p}(\yb\,|\,\thetab)$ and ${\bf p}(\yb, \xb\,|\,\thetab)$ 
denote the incomplete and the complete 
probability distributions respectively. \\

A fundamental property of the EM algorithm is the fact that it 
ensures the monotonous increasing  
of the incomplete likelihood function.
Any value of $\thetab$ such that $Q(\thetab, \thetab^{j}) \geq 
Q(\thetab^{j}, \thetab^{j})$ increases 
as well the incomplete log-likelihood, $i.e$, $\mathcal{L}(\thetab) 
\geq \mathcal{L}_{i}(\thetab^{j})$. 
Moreover, $\hat{\thetab}$ is a critical point of the incomplete 
likelihood ${\bf p}(\yb\,|\,\thetab)$ 
if and only if it is a fixed point of the re-estimation 
transformation, $\Mc$. 
A more detailed description of the convergence properties of the EM 
algorithm can be found in \cite{Dempster77}. \\
\\
%%%%%%%%%%%%%%%%%%%%%%%%%%%%%%%%%%%%%%%%%%%%%%%
\noindent {\bf Basic steps in the EM algorithm} 

The implementation of the EM algorithm starts by choosing an initial 
guess for the unknown
parameters, $\thetab^{0}$, which is used in the first iteration of 
the algorithm. Then, at each iteration
$j+1$ the following basic steps are performed 
\bit
\item {\bf i)} $\thetab=\wh{\thetab}^{j}$.
\item {\bf ii)} $\esp$-step: Compute $Q(\thetab',\thetab)$.
\item {\bf iii)} M-step: Find $\thetab^{j+1}$ such that 
$Q(\thetab^{j+1},\thetab) \geq Q(\thetab',\thetab)$ for all $\thetab' 
\in \Thetab$.
\eit
The iterative procedure is stopped when a fixed point is reached so 
that $\thetab^{j+1}=\thetab^{j+1}$.
\\
\\
%%%%%%%%%%%%%%%%%%%%%%%%%%%%%%%%%%%%%%%%%%%%
\noindent{\bf Penalized EM algorithm} 

In a Bayesian framework, we can consider the parameters $\thetab$ as  
random variables distributed according to an \aprio  ${\bf 
p}(\thetab\,|\,\xi)$. The  \apost distribution  for $\thetab$, ${\bf 
p}(\thetab\,|\,\yb,\, \xi)$ can be considered as a penalized version 
of the likelihood  ${\bf p}(\yb\,|\,\thetab)$ such that:
\[
{\bf p}(\thetab\,|\,\yb,\, \xi) \propto {\bf 
p}(\yb\,|\,\thetab)\,{\bf p}(\thetab\,|\,\xi)
\]
 The EM algorithm can be extended to a penalized version that 
converges to a local maximum of the penalized version of the 
likelihood function \cite{Hero85}. The penalized version of the EM 
functional  $Q_p(\thetab, \thetab^{k})$ is given by:
 \[\barr{lll}
 Q_p(\thetab, \thetab^{j})&=&\Esp{\log{{\bf p}(\yb, \xb,\,\thetab\,| 
\,\xi)}\,|\,\yb, \thetab^{j}} \\
 ~&=&\Esp{\log{{\bf p}(\yb, \xb\,|\,\thetab)}\,|\,\yb, \thetab^{j}}+ 
\log{{\bf p}(\thetab\,|\,\xi)}\\
 ~&=&Q_{ML}(\thetab, \thetab^{j})+\log{{\bf p}(\thetab\,|\,\xi)}
 \earr
 \]  
\subsection{Application of the EM-algorithm to source separation}
\noindent {\bf Missing information model}

Now we consider the application of the EM algorithm to the problem 
of source separation in Fourier space. For this reason, we call it {\bf Spectral EM} algorithm. 

As discussed in  section $2$, at each mode $\kb$, the vector of 
observations  $\yb$ arises from the following model
\[
\yb(\kb)=\Ab\,\sb(\kb)+\nb(\kb)
\]
Matrix $\Ab$ is the unknown (or partially unknown) mixing matrix and we assume that $\nb(\kb)$ is a 
zero mean Gaussian white noise with 
unknown diagonal covariance matrix $\Reps$. We will consider 
$\thetab=(\Ab,\Reps)$ the unknown
parameters to be estimated. 

The incomplete likelihood function we want to maximize, ${\bf 
p}(\yb_{1..K}\,|\,\Ab,\,\Reps)$, can be derived by marginalizing the 
joint distribution ${\bf p}(\yb_{1..K},\,\sb_{1..K}\,|\,\Ab,\,\Reps)$ over sources as follows:
\[
{\bf p}(\yb_{1..K}\,|\,\Ab,\,\Reps)=\int_{\sb_{1..K}} {\bf 
p}(\yb_{1..K},\,\sb_{1..K}\,|\,\Ab,\,\Reps)\,d\,\sb_{1..K}
\]

In our method, we assume a prior knowledge of the sources spatial power spectra, i.e.
the components $\sb(\kb)$ are assumed 
to follow an \aprio probability
distribution of the form
$${\bf p}(\sb(\kb)\,|\,\Cb(\kb))$$
where $(\Cb_k)_{k \in \Ib}$ represent the \aprio  spatial power 
spectrum shape. 

Assuming stationarity for the sources, 
the Fourier modes
are not correlated and we can perform the separation at each Fourier 
mode
independently.  The independence in the spectral domain can be 
written as:
\[\left\{\barr{lll}
{\bf p}\left( \,\,(\sb(\kb))_{ \kb \in \Ib} \,\,|\,\,(\Cb(\kb))_{ \kb 
\in \Ib}  \,\,\right)= \displaystyle{\prod_{\kb \in \Ib}{\bf 
p}(\sb(\kb)\,|\,\Cb(\kb))} \\~\\
{\bf p}\left( \,\,(\nb(\kb))_{ \kb \in \Ib} \,\,|\,\,\Reps  
\,\,\right)= \displaystyle{\prod_{\kb \in \Ib}{{\bf 
p}(\nb(\kb)\,|\,\Reps)}} \\
\earr \right.
\]
Because of this independence, $\kb$ is replaced  by the  integer 
index $k$ and $\Ib=\{1..K\}$ is a fixed arbitrary arrangement, where 
$K$ is the number of Fourier modes.

We can write  the complete data 
$(\yb_{1..K},\,\sb_{1..K})$ probability distribution as
\[
{\bf 
p}(\yb_{1..K},\,\sb_{1..K}\,|\,\Ab,\,\Reps)=\displaystyle{\prod_{k=1}^{K}}{\bf 
p}(\yb_k\,|\,\sb_k\,\Ab,\,\Reps)\,\displaystyle{\prod_{k=1}^{K}}{\bf 
p}(\sb_k\,|\,\Cb_k)
\]
so that,  the complete log-likelihood function $\mathcal{L}_c$ is 
given by
\beq\barr{lll}\label{eq:loglik}
\mathcal{L}_c(\Ab,\,\Reps)
&=&{\bf \log}\, {\bf p}(\yb_{1..K},\,\sb_{1..K}\,|\,\Ab,\,\Reps)\\~\\
~&=&\displaystyle{\sum_{k=1}^{K}}{\bf \log}\,{\bf 
p}(\yb_k\,|\,\sb_k\,\Ab,\,\Reps)\,+\,\displaystyle{\sum_{k=1}^{K}}{\bf \log}\,{\bf 
p}(\sb_k\,|\,\Cb_k)\\~\\
~&=&\displaystyle{\sum_{k=1}^{K}}\left( 
-\frac{1}{2}\,|\,2\,\pi\,\Reps\,|-\frac{1}{2}(\yb_k-\Ab\,\sb_k)^{T}\Reps^{-1}\,(\yb_k-\Ab\,\sb_k) 
\right)+cst
\earr
\eeq

\noindent{\bf Implementation of the Spectral EM algorithm} \\

The functional $Q(\thetab,\,\thetab^{k})$ is given by
\[\barr{lll}
Q \left( \thetab\,|\,\thetab^{j}\right) 
&=&\esp_{\sb\,|\,\thetab^{j}}\left[  
\log\,p(\yb_{1..K},\,\sb_{1..K}\,|\,\thetab) 
\,|\,\yb_{1..K},\,\thetab^{j} \right]\\~\\
~&=&\int_{\sb_{1..K}}\log\,p(\yb_{1..K},\,\sb_{1..K}\,|\,\thetab)\,p(\sb_{1..K}\,|\,\yb_{1..K},\,\thetab^{j})\,d\,\sb_{1..K}

\earr
\]

and using equation \ref{eq:loglik}, we get,
\[
Q \left( \thetab\,|\,\thetab^{j}\right) =   
-\frac{K}{2}\log{2\pi\Rb_{\epsilon}}-\frac{K}{2}\Tr{\Rb_{\epsilon}^{-1}(\wh{\Rb}_{yy}-
\Ab\wh{\Rb}_{ys}^{T}-\wh{\Rb}_{ys}\Ab^{T}+\Ab\wh{\Rb}_{ss}\Ab^{T})}+cst 
\] 
where
\beq\left\{ \barr{lll}\label{statistics}
\wh{\Rb}_{yy} &=& \frac {1}{K} \sum_{k}\yb_k\yb_k^{T}\\
\wh{\Rb}_{ys} &=& \frac {1}{K} \sum_{k}\yb_k\esp\,\left[\sb_k^{T} 
\,|\,\yb_k,\thetab^{j}\right]\\
\wh{\Rb}_{ss} &=& \frac {1}{K} \sum_{k}\esp\,\left[\sb_k\sb_k^{T} 
\,|\,\yb_k,\thetab^{j}\right]
 \earr
\right.
\eeq

To obtain the parameter $\thetab^{j+1}=(\Ab^{j+1}, \Reps^{j+1})$ at 
iteration $j+1$,  we  solve the  following gradient equations with 
respect to $\Ab$ and $\Reps$ in order to maximize the functional $Q$.
The complete log-likelihood function $\mathcal{L}_c$ of
equation \ref{eq:loglik} is quadratic as 
a function of $\Ab$ and consequently, the functional 
$Q(\Ab,\,\Reps\,|\,\Ab^{j},\,\Reps^{j})$  is also {\bf quadratic} in 
$\Ab$ and its  derivative  with respect to $\Reps$ can be easily 
calculated. Therefore,  $\wh{\Ab}^{j+1}$ and $\wh{\Reps}^{j+1}$ are 
easily derived.
\beq\left\{\barr{l}\label{systeme}
\frac{\partial{Q}}{\partial{\Ab}}=K\Rb_{\epsilon}^{-1}(\wh{\Rb}_{ys}-\Ab 
\wh{\Rb}_{ss})=0\\
~\\
\frac{\partial{Q}}{\partial{\Rb_{\epsilon}}}=\frac{K}{2}\Rb_{\epsilon}^{-1}(\wh{\Rb}_{yy}-\Ab\wh{\Rb}_{ys}^{T}-\wh{\Rb}_{ys}\Ab^{T}+\Ab\wh{\Rb}_{ss}\Ab^{T})\Rb_{\epsilon}^{-1}-\frac{K}{2}\Rb_{\epsilon}^{-1}=0
\earr \right.
\eeq
which yields to the re-estimation equations
\beq\left\{\barr{ll}\label{reestimation}
\Ab^{j+1}=\wh{\Rb}_{ys}\,(\wh{\Rb}_{ss})^{-1}\\~\\
\Reps^{j+1}=\wh{\Rb}_{yy}-\Ab^{j+1}\wh{\Rb}_{ys}^{T}-\wh{\Rb}_{ys}(\Ab^{j+1})^{T}+\Ab^{j+1}\wh{\Rb}_{ss}(\Ab^{j+1})^{T}
\earr\right.
\eeq

We now focus on the computation of the statistics (\ref{statistics})  
involved in the re-estimation transformation 
(\ref{reestimation}). The statistic $\wh{\Rb}_{yy}$, which represents 
the auto-covariance of the observations, is 
directly computed from the data $\yb_{1..K}$,  and remains constant 
throughout iterations of the EM algorithm. 
The statistic $\wh{\Rb}_{ys}$ represents the \apost  cross-covariance 
of the observations and the sources and the 
statistic $\wh{\Rb}_{ss}$ represents the \apost  auto-covariance of 
the sources. The  statistics $\wh{\Rb}_{ys}$ 
and $\wh{\Rb}_{ss}$ need, respectively, the computation of the \apost 
expectation 
$\esp \left[\sb_k\,|\,\yb_k,\,\thetab^{j} \right]$ and the \apost 
covariance 
$\esp \left[\sb_k\,\sb_k^{T}\,|\,\yb_k,\,\thetab^{j} \right]$ which 
depends on the  \aprio distribution 
of the sources ${\bf p}(\sb(\kb)\,|\,\Cb(\kb))$. We consider in the 
following two 
different prior distributions: \\
\\
{\bf (i)} Gaussian prior which corresponds to the Wiener filtering
\\ 
{\bf (ii)} Entropic prior which corresponds to the MEM deconvolution.\\
\\
\\
\noindent{\bf Gaussian prior:}

The \aprio distribution of the source vector $\sb_k$ is assumed to be 
a zero mean Gaussian with covariance $\Cb_k$:
\[
{\bf p}(\sb_k\,|\,\Cb_k) \propto     
\Exp{-\frac{1}{2}\sb_k^{T}\Cb_k^{-1}\sb_k}
\]
Then, the \apost distribution is
\[
p(\sb_k| \yb_k, \Ab, \Reps) \propto \Exp{-\Phi(\sb_k)} 
\]
with 
\[
\Phi(\sb_k)=\frac{1}{2} 
(\yb_k-\Ab\sb_k)^{T}\Reps^{-1}(\yb_k-\Ab\sb_k)+\frac{1}{2}\sb_k^{T}\Cb_k^{-1}\sb_k
\]
Thus, the expectations are easily derived as follows
\[\left\{\barr{ccccc}\label{gaussien}
\esp [\sb_k] =\eb_k = \left[ 
\Ab^{T}\Reps^{-1}\Ab+\Cb_k^{-1}\right]^{-1}\Ab^{T}\Reps^{-1}\yb_k \\
~\\
\esp [\sb_k\sb_k^{t}]  = \left[ 
\Ab^{T}\Reps^{-1}\Ab+\Cb_k^{-1}\right]^{-1}+\eb_k\eb_k^{T} 
\earr
\right.
\]

\vspace{1cm}
\noindent{\bf Entropic prior:}

For most foreground components the Gaussian prior does not fully 
represent the
underlying physical processes. For this reason, Hobson et al. 
\cite{1998MNRAS.300....1H}
introduced an entropic prior for the sources. 
Defining $\hb_k$ a  hidden vector such that $\sb_k=\Lb_k\hb_k$, where 
$\Lb_k$ 
is obtained from the  Cholesky decomposition of the spectrum of 
sources $\Cb_k$, 
the entropic prior can be expressed as follows
\[
p(\hb_k\,|\,\mb_k,\,\alpha) \propto \Exp{\alpha \Sc(\hb_k,\mb_k)} 
\]
where $\Sc(\hb_k,\mb_k)$ is the complex cross-entropy of $\hb_k$ and 
$\mb_k$ is a default model for $\hb_k$
\[\left\{\barr{lll}
\Sc(\hb_k,\mb_k)=\sum_{j=1}^{n}\left[\psi_{kj}-2 
m_{kj}-h_{kj}\log{\frac{\psi_{kj}+h_{kj}}{2 m_{kj}}}\right]\\
~\\
\psi_{kj}=\sqrt{h_{kj}^2+4 m_{kj}^{2}}\\
~\\
\mb_k=\left(\barr{l} 1\\ \vdots \\ 1 \earr \right) 

\earr
\right.
\]
Then, the  \apost distribution of $\hb_k$ is:  
\[
p(\hb_k| \yb_k, \Ab, \Reps) \propto \Exp{-\Phi(\hb_k)} 
\]
with
\[
\Phi(\hb_k)=\frac{1}{2} 
(\yb_k-\Ab\Lb\hb_k)^{T}\Reps^{-1}(\yb_k-\Ab\Lb\hb_k)-\alpha 
\Sc(\hb_k,\mb_k)
\]

The function $\Phi(\hb_k)$ being convex, the  \apost expectation of 
$\hb_k$ can be approximated by 
the minimum $\wh{\hb}_k$ of this function and the covariance by the 
inverse of the Hessian 
$[\Hb(\wh{\hb}_k)]^{-1}$ computed at this minimum:
\[\left\{\barr{l}
\esp [\sb_k]=\Lb \wh{\hb}_k\\
~\\
\esp [\sb_k\sb_k^{T}]=\Lb[\Hb(\wh{\hb}_k)]^{-1}\Lb^{T}+\esp 
[\sb_k]\esp [\sb_k]^{*}
\earr
\right.
\]

The minimum  $\wh{\hb}_k$ can be computed with an iterative 
minimization  algorithm. Expressions for the gradient and the  
hessian of $\Phi$ are easily derived,   
\[\left\{ \barr{l}
\nabla_{\hb_k} \Phi = \Lb^{T}\Ab^{T}\Reps^{-1}(\Ab\Lb\hb_k-\yb_k) +  
\alpha\,\left[\log{\left(\frac{\psi_{kj}+h_kj}{2mu_j}\right)}\right]_{j=1..n}\\

~\\
\nabla^{2}_{\hb_k} \Phi = \Lb^{T}\Ab^{T}\Reps^{-1}\Ab\Lb+ \alpha 
\,{\bf diag}\left[ 2m_kj/\psi_{kj} \right]
\earr
\right.
\]

\section{Application to Planck HFI simulated observations}

\subsection{Prior information}

The penalized form of the spectral EM algorithm permits to 
naturally include {\it a priori} information about the parameters we want
to estimate.  In the case of CMB 
observations, one will naturally dispose of additional information 
from theoretical predictions or from previous observations at the same 
or other electromagnetic frequencies (although in most cases at lower 
resolution).  We can take those into account in the separation 
procedure.  In particular, the electromagnetic 
spectrum of the anisotropies of the CMB radiation is accurately known at centimetre and 
millimetre wavelengths, and therefore there is no need to estimate it.  
Further, the emission from dust grains has also been measured by the 
IRAS satellite at 100 $\mu$m providing a first class template of it.  
\\

For a first test of the convergence capabilities of the algorithm we 
have only considered weak a priori information about the spatial 
spectrum of the components.  For the CMB component, the {\it a priori} 
spatial power spectrum was taken to be the theoretical model fitting 
best current observations \cite{jgpaper} and for dust, the mean spatial 
spectrum measured by the IRAS satellite at 100$\mu$m 
\cite{1998ApJ...500..525S}.  A circular spatial spectrum was 
assumed as a prior for the SZ component. This spectrum was obtained 
from an empirical fit to numerical simulations of SZ templates 
\cite{jjpaper}.  No prior information was assumed for the 
electromagnetic spectrum of the components, i.e., no prior on the 
mixing matrix.

\subsection{Results}

%%%%%%%%%%%%%%%%%%%%%%%%%%%%%%%%%%%%%%%%%%%%%%%%%%%%%%%%%%%%%%%%%%%%%
\begin{figure}
\label{recmaps}
\includegraphics[scale=0.73,draft=false]{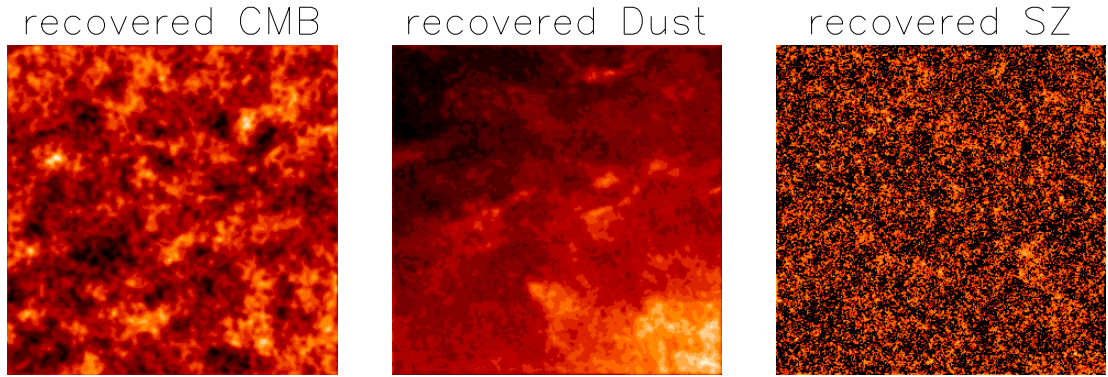}
\caption{Recovered spatial template of the CMB, dust and SZ 
components. For visibility, the recovered SZ is displayed in log
scale.}
\end{figure}
%%%%%%%%%%%%%%%%%%%%%%%%%%%%%%%%%%%%%%%%%%%%%%%%%%%%%%%%%%%%%%%%%%%%%

%%%%%%%%%%%%%%%%%%%%%%%%%%%%%%%%%%%%%%%%%%%%%%%%%%%%%%%%%%%%%%%%%%%%%
\begin{figure}
\label{recelecspec}
\includegraphics[scale=0.45,draft=false]{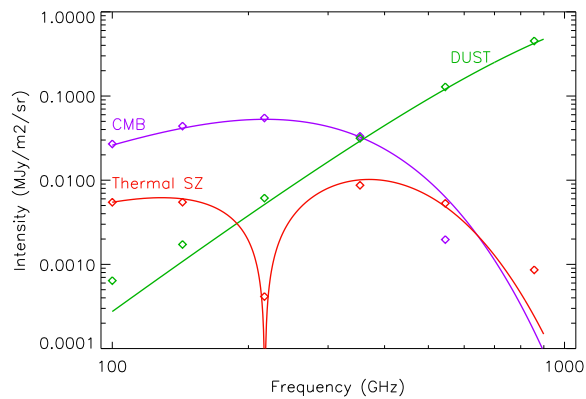}
\caption{The figure shows the recovered electromagnetic spectrum
of the components (diamonds) overplotted to the input spectrum
(solid lines).}
\end{figure}
%%%%%%%%%%%%%%%%%%%%%%%%%%%%%%%%%%%%%%%%%%%%%%%%%%%%%%%%%%%%%%%%%%%%%

The spectral EM algorithm has been applied to the simulated HFI 
observations presented in section 3.  The prior information on the 
spatial spectra of the components is used for a Wiener filtering 
estimation of the missing data at each iteration of the algorithm.  
The algorithm has been implemented in C and needs about 20000 
iterations to attain complete separation running on a 1GHz PENTIUM IV 
PC, for a total running time of about 8 hours of CPU time on $300 
\times 300$ pixels maps.  \\

The recovered spatial distributions of the components are shown in 
figure \ref{recmaps}.  CMB and dust components are recovered 
satisfactorily, with signal to noise ratio of about 5, better than the 
signal to noise ratio in any of the detectors.  The SZ is recovered 
with signal to noise of 1, not very satisfactorily (although the
brightest clusters can be picked by eye on the output).  This is 
essentially due to its peaky structure for which a Fourier treatment 
with very limited spectral {\it a priori} information is not very 
appropriate.  \\

The recovered electromagnetic spectra are shown in figure 
\ref{recelecspec}.  The CMB and dust spectra are recovered to better 
than 1\% for all detectors where the level of the components is not 
negligible.  The recovered electromagnetic spectrum of the SZ emission 
was multiplied by a global {\it calibration} factor due to a mismatch 
between the assumed power spectrum and the true (empirical) one.  In 
this way the true spectrum is recovered to better than 10\% for those 
channels with significant SZ contribution.

\section{Conclusions}

The spectral version of the EM algorithm presented in this paper 
constitutes a useful tool for the semi-blind separation of 
astrophysical components in noisy mixtures and in particular to 
separate CMB and foreground components in future space observations of 
the CMB. A first test of the algorithm on simulated Planck HFI 
observations with 3 components, CMB, dust and SZ emissions, has been 
proven successful.  We have been able to achieve complete separation 
with signal to noise of 5 and also to recover the electromagnetic 
spectrum of the components to better than 1 \% in about 20.000 
iterations.\\

In the present analysis we do not consider the smearing out of the 
observations due to the finite resolution of the detectors.  However, 
the spectral EM algorithm can be easily modified as proposed in 
section 2 to account for this.  A modified version of algorithm is 
already under testing and will permit the separation of 
multi-frequency observations at different resolutions.  \\
  
Further development of the algorithm is in progress including better 
use of the available prior information.  In real experiments, one 
would like to take advantage of the fact that the electromagnetic 
spectrum is known for some of the components in which case there is no 
need to estimate it within the separation algorithm.  The EM algorithm 
as presented here can naturally account for this fact by either 
reducing the number of free parameters to estimate or including a 
penalization term when only partial information is available.  A first 
version of the algorithm implementing the former is already available 
and produces good preliminary results.  \\

The major drawback of our present implementation of the spectral EM 
algorithm is the computation time needed for convergence (about 20000 
iterations and 8 hours of CPU to converge to complete separation).  
The acceleration of the algorithm is needed before it can be applied 
to full sky data sets or tested statistically by Monte-Carlo methods.  
Acceleration techniques are currently being
explored.  \\
{\small 

%\bibliographystyle{ieeeji}
%\bibliography{biben,revuedef,revueabr,baseAJ,baseKZ,gpipubli,bss}
\bibliographystyle{abbrvnat}
\bibliography{maxent03_astro_paper}
}  

\edoc